\documentstyle[preprint,aps]{revtex}
\tightenlines
\begin{document}
\draft
\title{Minimal extended flavor groups, matter fields chiral 
representations, and the flavor question}
\author{A. Doff and F. Pisano}
\address{Departamento de F\'\i sica, Universidade Federal do Paran\'a 
        81531--990 Curitiba, PR, Brazil}
\date
\today
\maketitle
\begin{abstract}
We show the specific unusual features on chiral gauge anomalies 
cancellation 
in the minimal, necessarily 3-3-1, and the largest 3-4-1 
weak isospin chiral gauge semisimple group leptoquark--bilepton 
extensions of the 3-2-1 conventional standard model of nuclear 
and electromagnetic interactions. In such models a natural explanation 
for the fundamental question of fermion generation replication arises 
from the self-consistency of a local gauge quantum field theory, which 
constrains the number of the QFD fermion families to the QCD color 
charges.
\end{abstract}
\bigskip
\bigskip
Flavor question has been a long standing puzzle. 
It was 
addressed in the SU(11) and SU(14) generalizations of the 
SU(5) grand unified theory with three and twelve families,~\cite{Georg} 
in the minimal supersymmetric model of quarks and leptons preon 
compositeness,~\cite{BabuPati} in the 3-3-1 leptoquark-bilepton 
model,~\cite{pp,Frampton92,Footetal,fp96}~\cite{Uedaetal} 
and 
in the Sp(6)$\times$U(1)$_Y$ 
symplectic gauge group extension.~\cite{Bagneid} 
The generation patterns of the reduced four-dimensional spinor 
fields with repeated copies of the same gauge quantum numbers is also a 
prediction of the 
5-dimensional space-time Kaluza--Klein theory.~\cite{KK} 
The anomaly cancellation of a gauge theory in all orders of 
perturbative expansion, which 
derives from the renormalizability condition, constrains the 
fermion representation content. Three perturbative anomalies have been 
identified~\cite{GeMa}
 for chiral gauge theories in four dimensional space-time: 
(i) The triangle chiral gauge anomaly~\cite{ABJ} must be cancelled to 
avoid violations of gauge invariance and the renormalizability of the theory; 
(ii) The global non-perturbative SU(2) chiral gauge anomaly,~\cite{Witten} 
which must be absent in order for the fermion integral to be defined 
in a gauge invariant way; (iii) 
The mixed perturbative chiral gauge gravitational 
anomaly~\cite{DelSalam,LAGW} which must be cancelled in order to 
ensure general covariance. In the standard model, the anomalies were 
cancelled into each generation of quarks and leptons. 
This is a very important issue, because in order to 
cancel anomalies we must sum over all hypercharges of quarks and 
leptons in each representation. 
In the standard model 
the anomaly-free conditions and flavor 
question do not seem to have any connection. This leads us to a question: 
Are the anomalies always cancelled automatically only by generation of 
quarks or leptons? And do the anomaly 
cancellation conditions have any connection with flavor puzzle? 
These questions of course cannot be answered within the standard model. 
It is of interest to find out some 
models in which these questions can be answered. 
In this note, we would like to provide some 
models in which the above questions can be answered. Particularly, 
we consider the anomaly cancellation and flavor question in the 3-3-1 
leptoquark-bilepton model and the 3-4-1 semisimple group generalization 
where there is a non-trivial very specific anomaly-free representation 
content.~\cite{pp,Frampton92,pp95} A bilepton is a boson which couples 
minimally with dimension four interactions to standard model leptons 
and can carry lepton number zero or 2.~\cite{Cuypers} 
The semisimple chiral gauge groups are 
\begin{equation}
G_{3n1}\equiv {\rm SU}(3)_c\otimes{\rm SU}(n)_L\otimes{\rm U(1)}_N
\label{ggttu}
\end{equation}
with $n=3,4$, from now on 3-$n$-1 models and $N$ is the new U(1)$_N$ 
charge of the Abelian semisimple group factor. For the standard model 
weak isospin factor, $n=2$. 
\par
The general anomaly-free condition is 
\begin{equation}
\sum_{\rm representations} {\rm Tr}[\{T^a_L, T^b_L\}T^c_L - \{T^a_R, T^b_R\}
T^c_R] = 0
\label{seilah}
\end{equation}
where $T^a$ are the gauge group generators and the $a$ index runs over 
the dimension of the simple SU($n$) group, $a= 1,2,...,n^2-1$, with a 
rank $n-1$, and $a=0$ for the Abelian factor.  
The standard model independent families have the following 
structure under the chiral 3-2-1 gauge group, 
\begin{mathletters}
\begin{equation}
f_{iL}\sim ({\bf 1}_c, {\bf 2}_L, Y=-1), \quad 
\ell_{iR}\sim ({\bf 1}_c, {\bf 1}_R, -2)
\label{lpsm}
\end{equation}
for the leptonic sector with the flavor labels $\ell_i = e,\mu,\tau$ 
($i = 1,2,3$) and 
\begin{equation}
Q_{iL}\sim ({\bf 3}_c, {\bf 2}_L, 1/3),\,\,
u_{iR}\sim ({\bf 3}_c, {\bf 1}_R, 4/3),\,\,
d_{iR}\sim ({\bf 3}_c, {\bf 1}_R, -2/3)
\label{qrks}
\end{equation}
\end{mathletters}
for the quark flavors. These symmetry eigenstates are related to the 
mass eigenstates by three Cabibbo--Kobayashi--Maskawa rotation angles. 
The electric charge operator is defined as 
\begin{equation}\frac{{\cal Q}}{e} = T_3 + T_0 = T_3 + \frac{Y}{2}.
\label{opch}
\end{equation}
with $e = g\sin\theta_W = g^\prime\cos\theta_W = 1.602\,177\,33(49)
\times 10^{-19}\,\,{\rm C} = 4.803\,206\,8(15)\times 10^{-8}\,\,
{\rm esu}$.~\cite{pdg} 
The matrices $T^a$ in Eq.~(\ref{seilah}) will be either a half of the 
Pauli matrices, $\tau^a$, $a=1,2,2^2-1$, or the 
weak hypercharge generator $T^0\equiv Y/2$ for $a=0$ proportional 
to the $n\times n$ identity operator. 
Since the weak isospin 
group SU(2) is a safe group,~\cite{Georgi72} then 
\begin{equation} 
{\rm Tr}(\{\tau^a, \tau^b\}\tau^c) = 2\delta^{ab}{\rm Tr}(\tau^c) = 0\,,
\label{acudt}
\end{equation} 
but in the case where at least one of the $T$ generators is the hypercharge 
$Y$, we have 
\begin{equation}
{\rm Tr}(\{\tau^a, \tau^b\}Y) = 2\delta^{ab}\,{\rm Tr}(Y)
\label{cinque}
\end{equation}
and 
\begin{equation}
{\rm Tr}(\tau^a Y Y)\propto {\rm Tr}(\tau^a).
\label{sei}
\end{equation}
The anomaly contribution in Eq.~(\ref{cinque}) is proportional 
to the sum of all fermionic discrete hypercharge values 
\begin{equation} 
{\rm Tr}(Y) = \sum_{\rm lepton} (Y_L + Y_R) + \sum_{\rm quark} (Y_L + Y_R).
\label{sette}
\end{equation}
The Tr$(Y)$ in Eq.~(\ref{sette}) vanishes for the fermion content in the 
equations~(\ref{lpsm}) and (\ref{qrks}), 
\begin{mathletters}
\begin{equation}
\sum_{\rm lepton}(Y_L + Y_R) = 
Y(\nu_{\ell L}) + Y(\ell_L) + Y(\ell_R) = 
2Y(f_{iL}) + Y(\ell_R) = -4
\label{ottopri}
\end{equation}
\begin{equation}
\sum_{\rm quark} (Y_L + Y_R) = 3[2Y(Q_{iL}) + Y(u_{iR}) + Y(d_{iR})] = +4
\label{ottosec}
\end{equation}
\end{mathletters}
where the global factor in the last equation takes into account the 
number of SU(3)$_c$ color charges. The summations indicated on 
the left-handed side involve the color, flavor, and the weak hypercharge 
degrees of freedom. 
For the case in which all $T$'s are 
the hypercharge U(1)$_Y$ generator, we have from Eq.~(\ref{opch}) 
\begin{equation}
{\rm Tr}(Y^3)\propto {\rm Tr}({\cal Q}^2T_3 - {\cal Q}T_3^2)
\label{dieci}
\end{equation}
where we used the fact that the electromagnetic vector 
neutral current vertices do not have anomalies. Thus the right-handed 
side of Eq.~(\ref{dieci}) yields to 
\begin{mathletters}
\begin{equation}
\sum_{\rm lepton} ({\cal Q}^2T_3 - {\cal Q}T^2_3) = 
[(0)^2(1/2) - (0)(1/2)^2] + [(-1)^2(-1/2) - (-1)(-1/2)^2] = 
-\frac{1}{4}
\label{undici}
\end{equation}
and 
\begin{eqnarray}
\sum_{\rm quark} ({\cal Q}^2T_3 - {\cal Q}T_3^2) & = & \nonumber \\ 
3[(2/3)^2(1/2) & - & (2/3)(1/2)^2] + 3[(-1/3)^2(-1/2) - (-1/3)(-1/2)^2] 
= + \frac{1}{4}
\label{dodici}
\end{eqnarray}
\end{mathletters}
which guarantees that the standard fermion assignments turn the 3-2-1 
standard theory chiral anomaly free. Taking into account the electric 
charge operator, the trace of an odd number of $Y$ generator in 
Eq.~(\ref{cinque}) and Eq.~(\ref{dieci}) can be re-written as follows:
\begin{equation}
{\rm Tr}(Y)\,\,\,{\rm or}\,\,\,{\rm Tr}(Y^3)\propto {\rm Tr}({\cal Q}) 
= \sum_j{\cal Q}_j.
\label{tredici}
\end{equation}
Two remarks can be considered here. The first one concerns the fact that 
in the standard model we have anomaly cancellation within each 
independent family of chiral fermions and not by generations. The 
other one is that the anomaly cancellation occurs in each chiral 
sector. 
\par
Now, let us consider the following 3-3-1 non-universal fermion 
representation content 
\begin{mathletters}
\begin{eqnarray}
f_{1L} \sim ({\bf 1}_c,{\bf 3}_L,0)&,& \nonumber \\ 
Q_{1L} \sim ({\bf 3}_c,{\bf 3}_L,2/3),\,\, 
u_{1R} \sim ({\bf 3}_c,{\bf 1}_R,2/3)&,& \\
\label{quatord}
d_{1R} \sim ({\bf 3}_c,{\bf 1}_R,-1/3),\,\, 
J_{1R} \sim ({\bf 3}_c,{\bf 1}_R,5/3), \nonumber
\end{eqnarray}
for the first generation and 
\begin{eqnarray}
f_{iL} \sim ({\bf 1}_c,{\bf 3}_L,0)&,& \nonumber \\
Q_{iL} \sim ({\bf 3}_c,\bar{\bf 3}_L,-1/3),\,\,
J_{iR} \sim ({\bf 3}_c,{\bf 1}_R,-4/3)&,& \\
\label{quindic}
u_{iR} \sim ({\bf 3}_c,{\bf 1}_R,2/3),\,\,
d_{iR} \sim ({\bf 3}_c,{\bf 1}_R,-1/3), \nonumber
\end{eqnarray}
\end{mathletters}
when $i=2,3$ for the second and third generations. 
The {\rm N}$\times${\rm N} flavor mixing matrix is paremetrized 
by ${\rm N}({\rm N}-1)/2$ rotation angles and $({\rm N}-1)({\rm N}-2)/2$ 
phases. 
Three generations of 
leptons, $f_{iL}$, $i=1,2,3$ transform in the same way, and 
right-handed flavor singlet fields are not introduced. 
In the electroweak semisimple group factor of the 3-3-1 gauge symmetry 
the electric charge operator is defined as 
\begin{equation}
\frac{{\cal Q}}{e} = \frac{1}{2}(\lambda_3 - \sqrt 3\lambda_8)+N
\label{opcr}
\end{equation}
where $\lambda_{3,8}$ are diagonal Gell-Mann matrices associated to 
the rank two of the SU(3)$_L$ flavor simple group and $N$ is the new 
U(1)$_N$ hypercharge, 
\begin{equation}
N = \frac{\sqrt 3}{2}\lambda_8 + \frac{Y}{2}
\label{ateqfm}
\end{equation}
equivalent to the mean electric charge of the fields contained in the 
multiplet. 
The electron electric charge in terms of the $\theta_{3-1}$ mixing 
angle of the SU(3)$_L$ and U(1)$_N$ simple and semisimple groups with 
the associated $g$ and $g^\prime$ gauge coupling constants is 
\begin{equation}
e^2 = \frac{g^2\sin^2\theta_{3-1}}{1 + 3\sin^2\theta_{3-1}} = 
\frac{g^{\prime 2}\cos^2\theta_{3-1}}{1 + 3\sin^2\theta_{3-1}}
\label{caelmt}
\end{equation}
and with $e^2 = g^2\sin^2\theta_W = g^{\prime 2} (1-4\sin^2\theta_W)$ the 
Eq.~(\ref{caelmt}) becomes 
\begin{equation}
\tan^2\theta_{3-1} = \frac{\sin^2\theta_W}{1 - 4\sin^2\theta_W}
\label{dptr}
\end{equation}
with the bound on the Weinberg electroweak angle $\sin^2\theta_W < 1/4$. 
The SU($n$) groups for $n > 2$ are not safe in the sense of 
Eq.~(\ref{acudt}). For $n=3$ the generators are proportional to 
the Gell-Mann matrices, closing among them the Lie algebra 
structure, 
\begin{mathletters}
\begin{eqnarray}
[\lambda_a,\,\lambda_b] & = & 2if_{abc}\lambda_c\, \\
\label{ddnove}
\{\lambda_a,\,\lambda_b\} & = & \frac{4}{3}\delta_{ab} + 2d_{abc}\lambda_c,
\label{dctren}
\end{eqnarray}
\end{mathletters}
where the structure constants $f_{abc}$ are totally antisymmetric 
and $d_{abc}$ are totally symmetric under the exchange of indices. 
For any $n\times n$ $\lambda$-matrices with arbitrary $n$, 
\begin{mathletters} 
\begin{eqnarray}
f_{abc} & = &\frac{1}{4i}\,{\rm Tr}([\lambda_a, \lambda_b]\lambda_c)\, \\
d_{abc} & = &\frac{1}{4}\,{\rm Tr}(\{\lambda_a, \lambda_b\}\lambda_c)\,,
\end{eqnarray}
\end{mathletters}
in general the anomaly is proportional to $d_{abc}$. The $d_{abc}$ 
constants vanish for $n=2$ case. When $n > 2$ the generalization of 
the anti-commutation relation in Eq.~(\ref{acudt}) is 
\begin{equation} 
\{\lambda_a, \lambda_b\} = \frac{4}{n}\delta_{ab} + 2d_{abc}\lambda_c\,.
\label{genctt}
\end{equation}
The anomaly-free conditions are
\begin{mathletters}
\begin{eqnarray}
{\rm Tr}([{\rm SU(3)}_c]^2[{\rm U(1)}_N]) & = & 0 \\
\label{unu1}
{\rm Tr}([{\rm SU(3)}_L]^2[{\rm U(1)}_N]) & = & 0 \\ 
\label{dusu}
{\rm Tr}([{\rm U(1)}_N]^3) & = & 0 \\
\label{trese}
{\rm Tr}([{\rm graviton}]^2[{\rm U(1)}_N]) & = & 0.
\end{eqnarray}
\end{mathletters}
Each generation is anomalous. Let us take the condition in Eq.~(\ref{dusu}) 
containing only the Abelian factor. The general cancellation scheme 
can be found in Ref.~\cite{fp96}. 
We can see that the $N^3$ anomaly of the first generation is 
\begin{eqnarray}
\sum_{\rm lepton} N_L & + & \sum_{\rm quark} (N_L + N_R) = \nonumber \\
0 + 3\,[(2/3)\times 3 & + & 5/3 + 2/3 + (-1/3)] = +12, 
\label{diciasete}
\end{eqnarray}
and the anomalies of the second or the third generations are
\begin{eqnarray}
\sum_{\rm lepton} N_L & + & \sum_{\rm quark} (N_L + N_R) = \nonumber \\
0 + 3\,[(-1/3) & + & (-4/3) + 2/3 + (-1/3)] = -6.
\label{diciotto}
\end{eqnarray}
It is clear that anomalies are not cancelled within each generation as in 
the standard model. The global factor 3 takes into 
account the SU(3)$_c$ three color charges. The general condition that, 
for a given fermionic representation ${\bf D}$ it holds~\cite{G84} 
\begin{equation}
{\cal A}({\bf D}) = -{\cal A}({\bar{\bf D}})
\label{drps}
\end{equation}
where ${\cal A}({\bar{\bf D}})$ is the anomaly of the conjugate 
representation of ${\bf D}$, and the fact that, as can be seen from 
Eqs.~(13), there are the same 
number of multiplets in ${\bf 3}$ and $\bar{\bf 3}$ representation 
turn the theory anomaly-free.  The anomaly cancellation in this case 
does not occur by chiral sector. 
\par
Now let us display a second representation content which contains 
heavy leptons.~\cite{PlTon93} For the first generation 
\begin{mathletters}
\begin{eqnarray}
f_{1L} \sim ({\bf 1}_c,{\bf 3}_L,0),\,\,
E_{1R} \sim ({\bf 1}_c,{\bf 1}_R,1),\,\,
\ell_{1R} \sim ({\bf 1}_c,{\bf 1}_R,-1)&,& \nonumber \\
Q_{1L} \sim ({\bf 3}_c,{\bf 3}_L,2/3),\,\,
u_{1R} \sim ({\bf 3}_c,{\bf 1}_R,2/3)&,& \\
\label{iijk}
d_{1R} \sim ({\bf 3}_c,{\bf 1}_R,-1/3),\,\,
J_{1R} \sim ({\bf 3}_c,{\bf 1}_R,5/3), \nonumber
\end{eqnarray}
and 
\begin{eqnarray}
f_{iL} \sim ({\bf 1}_c,{\bf 3}_L,0),\,\,
E_{iR} \sim ({\bf 1}_c,{\bf 1}_R, 1),\,\,
\ell_{iR} \sim ({\bf 1}_c,{\bf 1}_R,-1)&,& \nonumber \\
Q_{iL} \sim ({\bf 3}_c,{\bar{\bf 3}}_L,-1/3),\,\,
u_{iR} \sim ({\bf 3}_c,{\bf 1}_R,2/3)&,& \\
\label{ikkr}
d_{iR} \sim ({\bf 3}_c,{\bf 1}_R,-1/3),\,\,
J_{iR} \sim ({\bf 3}_c,{\bf 1}_R,-4/3), \nonumber
\end{eqnarray}
\end{mathletters}
for the second and third generations, $i=2,3$. The $N^3$ 
anomaly of the first generation is
\begin{mathletters}
\begin{eqnarray}
\sum_{\rm lepton} (N_L + N_R) & +&  \sum_{\rm quark} (N_L + N_R) = 
\nonumber \\
0\times 3 + (-1) + 1 + 3\,[(2/3)\times 3 & + &  (5/3) + (2/3) + (-1/3)] = 
+12
\label{vetidue}
\end{eqnarray}
and
\begin{eqnarray}
\sum_{\rm lepton}(N_L + N_R) & + & \sum_{\rm quark} (N_L + N_R) = \nonumber \\
0\times 3 + (-1) + 1 + 3\,[(-1/3)\times 3 & + &  (-4/3) + (2/3) + (-1/3)] = 
-6
\label{vetitre}
\end{eqnarray}
\end{mathletters}
are the anomalies of the second and the third generations and the 
complete generations $N^3$ anomaly vanishes again. 
\par
Let us now consider the following 3-3-1 fermionic assignments for the 
first and second generations~\cite{Frampton92} 
\begin{mathletters}
\begin{eqnarray}
f_{iL} \sim ({\bf 1}_c,{\bar{\bf 3}}_L,-1/3),\,\, 
\ell_{iR} \sim ({\bf 1}_c,{\bf 1}_R,-1)&,& \nonumber \\
Q_{iL} \sim ({\bf 3}_c,{\bf 3}_L,0), \,\,
u_{iR} \sim ({\bf 3}_c,{\bf 1}_R,2/3), \\
d_{iR} \sim ({\bf 3}_c,{\bf 1}_R,-1/3),\,\,
d^\prime_{iR} \sim ({\bf 1}_c,{\bf 1}_R,-1/3), \nonumber
\end{eqnarray}
$i=1,2$ and 
\begin{eqnarray}
f_{3L} \sim ({\bf 1}_c,\bar{\bf 3}_L,-1/3),\,\,
\ell_{3R} \sim ({\bf 1}_c,{\bf 1}_R,-1)&,& \nonumber \\
Q_{3L} \sim ({\bf 3}_c,\bar{\bf 3}_L,1/3),\,\,
u_{3R} \sim ({\bf 3}_c,{\bf 1}_R,2/3), \\
u^\prime_{3R} \sim ({\bf 3}_c,{\bf 1}_R,2/3),\,\,
d_{3R} \sim ({\bf 3}_c,{\bf 1}_R,-1/3), \nonumber
\end{eqnarray}
\end{mathletters}
for the third generation. Such peculiar representation structure 
contains the same number of flavor triplets and anti-triplets. 
The anomaly of the first or the second generation is 
\begin{eqnarray}
\sum_{\rm lepton} (N_L + N_R) & + & \sum_{\rm quark} (N_L + N_R) = \nonumber \\
(-1/3)\times 3 + (-1) + 3\,[(0)\times 3 & + & (2/3) + (-1/3) + (-1/3)] = -2
\label{vesei}
\end{eqnarray}
and for the third generation
\begin{eqnarray}
\sum_{\rm lepton} (N_L + N_R) & + & \sum_{\rm quark} (N_L + N_R) = \nonumber \\
(-1/3)\times 3 + (-1) + 3\,[(1/3)\times 3 & + & (2/3) + (2/3) + (-1/3)] = 
+4\,,
\label{vensete}
\end{eqnarray}
so the $N^3$ anomalies cancel. Note that we do not have anomaly cancellation 
by chiral sector. 
\par
Finally, we consider the following 3-4-1 fermionic representation 
content.~\cite{pp95} 
In the first generation we have a ${\bf 4}$-plet of leptons
\begin{mathletters}
\begin{equation}
f_{1L} \sim ({\bf 1}_c,{\bf 4}_L,0)
\label{venotto}
\end{equation}
which contains all light leptonic degrees of freedom so that we do not 
need to introduce right-handed singlet fields. The assignments of the 
quarks are 
\begin{eqnarray}
Q_{1L} \sim ({\bf 3}_c,{\bf 4}_L,2/3),\,\, 
u_{1R} \sim ({\bf 3}_c,{\bf 1}_R,2/3),\,\,
d_{1R} \sim ({\bf 3}_c,{\bf 1}_R,-1/3)&,& \nonumber \\
u^\prime_{1R} \sim ({\bf 3}_c,{\bf 1}_R,2/3),\,\,
J_{1R} \sim ({\bf 3}_c,{\bf 1}_R, 5/3)&.&
\label{vennove}
\end{eqnarray}
and the last two families have the following fundamental representation 
transformation properties under the 3-4-1 gauge group 
\begin{eqnarray}
f_{iL} \sim ({\bf 1}_c,{\bf  4}_L,0)&,& \nonumber \\
Q_{iL} \sim ({\bf 3}_c,\bar{\bf 4}_L,-1/3),\,\,
J_{iR} \sim ({\bf 3}_c,{\bf 1}_R,-4/3)&,& \\
d^\prime_{iR} \sim ({\bf 3}_c,{\bf 1}_R,-1/3),\,\,
u_{iR} \sim ({\bf 3}_c,{\bf 1}_R,2/3),\,\,
d_{iR} \sim ({\bf 3}_c,{\bf 1}_R,1/3)&.& \nonumber
\end{eqnarray}
\end{mathletters}
In this extension we enlarge the electric charge operator to 
\begin{equation}
\frac{{\cal Q}}{e} = \frac{1}{2} (\lambda_3 + \frac{1}{\sqrt 3} 
\lambda_8 - \frac{2\sqrt 6}{3}\lambda_{15}) + N^\prime
\label{opcarga}
\end{equation}
where $\lambda_{3,8,15}$ are diagonal $4\times 4$ matrices whose 
number is equivalent to the rank of the SU(4) simple group. 
In the same way as the 3-3-1 
fermion representations we can verify that the anomaly cancellation 
occurs between generations. The anomaly of the first generation is 
\begin{mathletters}
\begin{eqnarray}
\sum_{\rm lepton} N^\prime_L & + & \sum_{\rm quark} 
(N^\prime_L + N^\prime_R) = \nonumber \\
0 + 3\,[(2/3)\times 3 + (2/3) & + & (-1/3) + (2/3) + (5/3)] = +14
\label{trdue}
\end{eqnarray}
and for each of the last two generations
\begin{eqnarray}
\sum_{\rm lepton} N^\prime_L & + & \sum_{\rm quark} 
(N^\prime_L + N^\prime_R) = \nonumber \\
0 + 3\,[(-1/3)\times 3 + (-4/3) & + & (-1/3) + (2/3) + (-1/3)] = -7
\label{trtr}
\end{eqnarray}
\end{mathletters}
so that all $N^3$ Abelian contributions cancel. 
\par
It should be emphasized that there is 
the same number of triplets 
(${\bf 4}$-plets) and antitriplets ($\bar{\bf 4}$-plets) if each quark 
flavor appears with three color charges, so that the non-Abelian 
anomalies cancel in all these cases. Contrary to the massive neutrino 
extensions~\cite{Mohapatra90} of the conventional standard model 
in the 3-3-1 and 3-4-1 leptoquark-bilepton 
models the electric charge quantization is unavoidable and does not 
depend on the Dirac, Majorana or Dirac--Majorana~\cite{Esposito} 
character of the neutral fermions.~\cite{Dofffp} There is always a  
set of leptonic generations transforming as 
$({\bf 1}_c, {\bf 3}_L,0)$ or as $({\bf 1}_c,{\bf 4}_L,0)$ 
with $N = N^\prime = 0$ 
and the electric charge operator is a 
linear combination of the weak isospin simple group generators whose 
number is the rank of the group. 
In the leptonic sector there is the fundamental representation content 
structure of grand unified theories. 
\par
This sort of minimal gauge semisimple group models has became an 
interesting possibility not only 
for an extension of the standard model but alternative theoretical 
possibilities at near energy scales below a few TeV 
answering several open fundamental questions such as 
the flavor question and addressing new physics close to the Fermi 
scale.~\cite{revetc,tutto,DuongMa} The flavor question has no 
explanation within the 3-2-1 standard model. The anomaly 
cancellation in the 3-$n$-1 leptoquark-bilepton models with 
$n=3,4$ is possible only if the number of generations of quarks and 
leptons is equal to the number of color charges. In general, these 
theories will be anomaly-free if the number of generations is a 
multiple of the number of colors with the lepton families having 
identical transformation properties with $N=0$ and without 
right-handed singlets even for massive neutrinos. 
In fact, the lightest known particles 
could be the sector in which a symmetry is uncovered, then the 
lepton sector is the part of the model determining new approximate 
symmetries. There is a close connection of the fundamental fermions 
generations, the three color charges, and the self-consistency of a 
local gauge quantum field theory. 
It is possible to answer the flavor question offering 
a connection between the QFD and QCD which is lost in the 
standard model. 
In these models the Weinberg angle has a Landau pole and is bounded 
from above giving an intrinsic limitation on the range of validity 
of the theory. This is another issue which does not have an equivalent 
one in the standard model. 
\acknowledgments
We are very grateful to Tuan A. Tran for their 
kindness and collaboration. A. D. wish to thank the CAPES (Brazil) 
for full financial support.

\end{document}